\documentstyle{aipproc}

\begin{document}
\title{Understanding Blazar Jets \\ Through Their Multifrequency Emission}

\author{Rita M. Sambruna$^*$}
\address{$^*$Pennsylvania State University\\
Dept. of Astronomy \& Astrophys\\
525 Davey Lab, University Park, PA 16802 \\
(rms@astro.psu.edu)} 

\maketitle

\input{psfig.tex}

\begin{abstract}

Being dominated by non-thermal (synchrotron and inverse Compton)
emission from a relativistic jet, blazars offer important clues to the
structure and radiative processes in extragalactic jets.  Crucial
information is provided by blazars' spectral energy distributions from
radio to gamma-rays (GeV and TeV energies), their trends with
bolometric luminosity, and their correlated variability
properties. This review is focussed on recent multiwavelength
monitorings of confirmed and candidate TeV blazars and the constraints
they provide for the radiative properties of the emitting particles. I
also present recent observations of the newly discovered class of
``blue quasars'' and the implications for current blazars' unification
schemes. 

\end{abstract}

\section*{The blazar family} 

Blazars are radio-loud Active Galactic Nuclei (AGNs) with polarized,
luminous, and rapidly variable non-thermal continuum emission,
extending from radio to gamma-rays (GeV and TeV energies), from a
relativistic jet oriented close to the line of sight. As such, they
are rare laboratories to study the physics and structure of
extragalactic jets, present in all radio-loud AGNs \cite{megunif}.
 
\begin{figure}
\noindent{\psfig{file=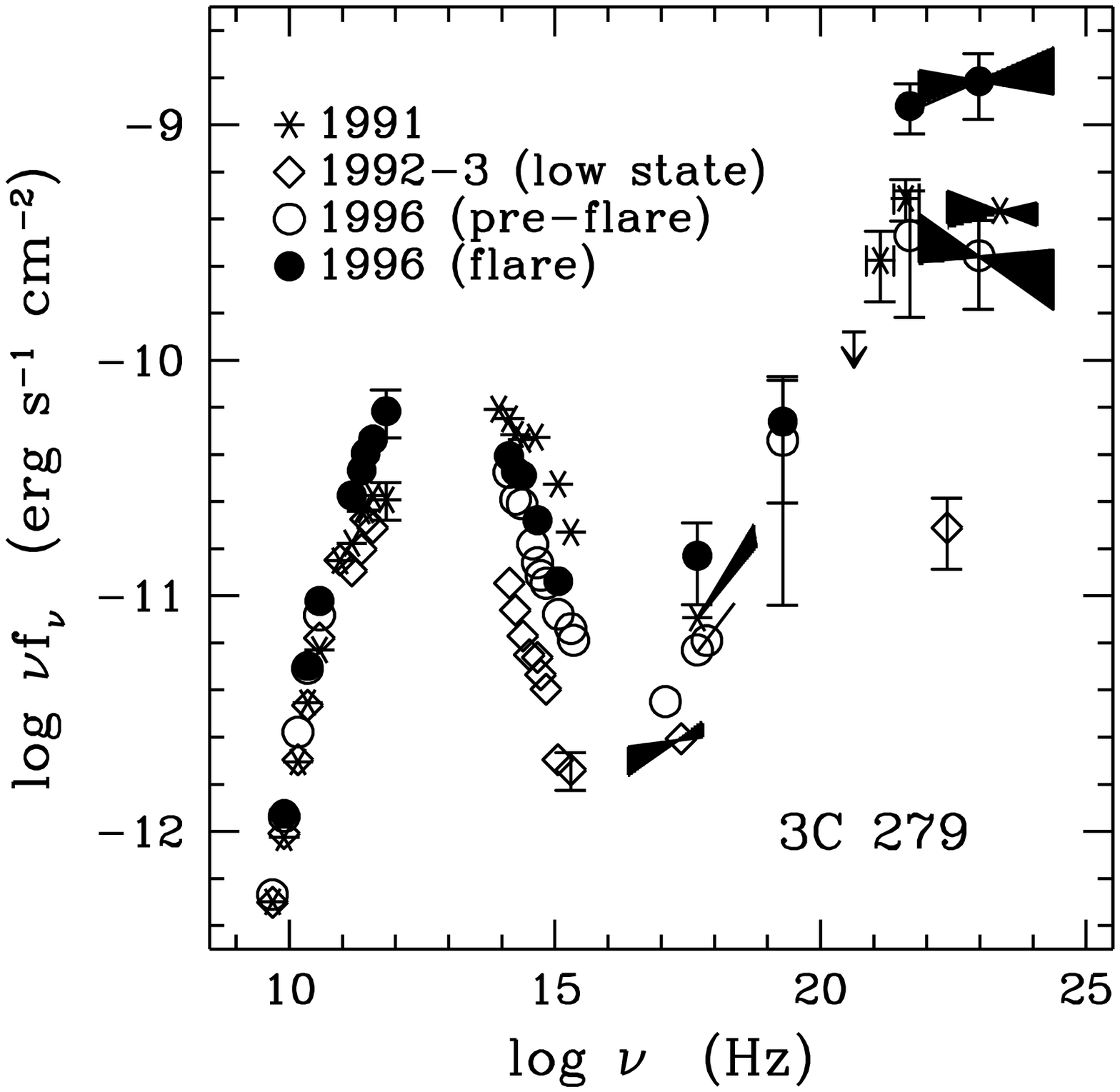,height=2.4in}}{\psfig{file=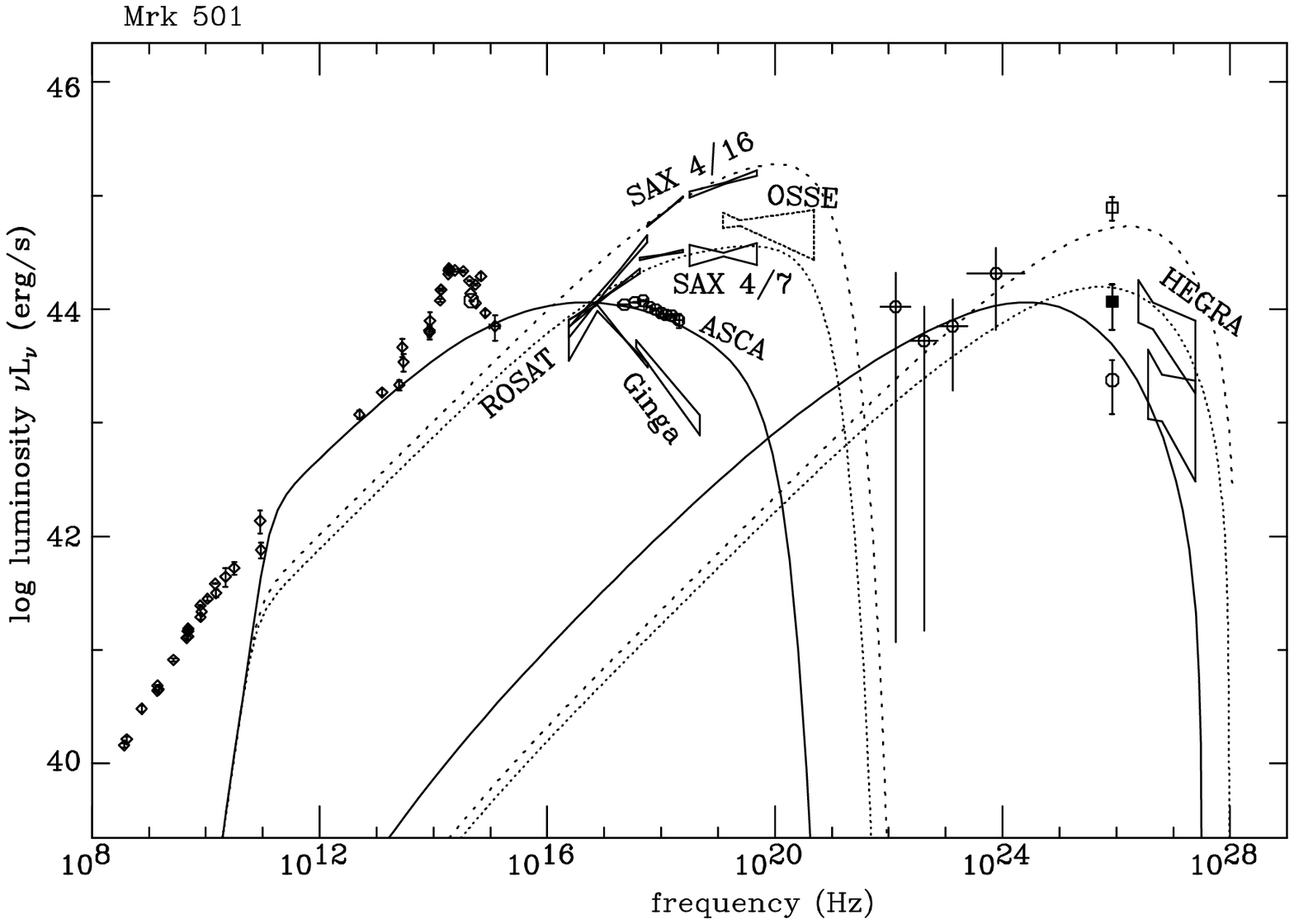,height=2.4in}}
\vspace{10pt}
\caption{Spectral energy distributions (SEDs) of the red [Left, (a)]
and blue [Right, (b):] blazars 3C 279 and Mrk 501, respectively (from
[2,3]).  Blazars' SEDs typically have two broad humps, the first
peaking anywhere from IR/optical (in red blazars like 3C279) to hard
X-rays (in blue blazars like Mrk 501) and due to synchrotron emission
from a relativistic jet.  The second component, extending to
gamma-rays, is less well understood. A popular explanation is inverse
Compton scattering of ambient seed photons off the jet's
electrons. The largest variability amplitudes are observed above the
peaks in both sources.}
\label{seds}
\end{figure}

Strong clues are provided by blazars' spectral energy distributions
(SEDs). These are typically double-humped (Fig. \ref{seds}; from
\cite{wehrle98,kataoka99a}), with the first component peaking at
IR/optical wavelengths in ``red blazars'' (also called Low-energy
peaked blazars, LBLs) and at UV/X-rays in ``blue blazars'' (or HBLs,
High-energy peaked blazars) \footnote[1]{A practical way to
discriminate between LBLs and HBLs is though their radio-to-X-ray
spectral indices, $\alpha_{rx}$, with $\alpha_{rx} > 0.8$ in LBLs and
$\alpha_{rx} < 0.8$ in HBLs \cite{padovani95}. Flat Spectrum Radio
Quasars (FSRQs) have SEDs similar to LBLs, but are more luminous and
have stronger optical emission lines \cite{scarpa}.}. Its rapid
variability and high polarization leave little doubt that it is due to
synchrotron emission from relativistic electrons in the jet. The
second spectral component extends from X-rays to gamma-rays, and its
origin is less well understood. In the leptonic models, it could be
due to inverse Compton (IC) scattering off the electrons of photons
either internal or external to the jet (synchrotron-self Compton, SSC
and external Compton, EC, respectively; see, e.g.,
\cite{bottcher99}). Here I will assume the leptonic models, but
acknowledge that an alternative is provided by the hadronic scenario
(e.g., \cite{rachen99}). 
 
Red and blue blazars are just the extrema of a continuous distribution
of SEDs \cite{rita96,rita97}. Indeed, deep multicolor surveys
\cite{perlman99,sally} are finding an increasing number of sources
with intermediate SED shapes, and new trends with jet bolometric
luminosity are discovered \cite{rita96,fossati98}. Specifically, the
lower-luminosity blue blazars have higher synchrotron and IC peak
frequencies, lower ratios of the IC to synchrotron peak fluxes, and
weaker or absent optical emission lines than their more luminous red
counterparts. 

A possible interpretation is that the different SEDs are due to
different predominant electrons' cooling mechanisms
\cite{ghisello98}. In a homogeneous scenario, the synchrotron peak
frequency $\nu_{peak} \propto \gamma_{peak}^2$, where $\gamma_{peak}$
is the electron energy determined by the competition between
acceleration and cooling. Because of the lower energy densities, in
line-less blue blazars the balance between heating and cooling is
achieved at larger $\gamma_{peak}$. In contrast, in red blazars the
electrons are more efficiently cooled due to the additional EC
component and reach a lower final $\gamma_{peak}$. The emerging
scenario is that blue blazars are SSC-dominated, while the EC
mechanism dominates the production of gamma-rays in red blazars.
While there are a few caveats to this picture \cite{meg99}, the clear
message is that the spectral diversity of blazars' jets cannot be
explained by beaming/orientation effects {\it only}, but require
instead a change of physical parameters and/or a different jet
environment \cite{rita96,markos99}. 

\section*{Probing blazars' paradigm: \\ 
Multiwavelength variability \\
of TeV blazars} 

Correlated multiwavelength variability provides a way to test the
cooling paradigm since the various synchrotron and IC models make
different predictions for the relative flare amplitudes and shape, and
the time lags. Since the same population of electrons is responsible
for emitting both spectral components (in a homogeneous scenario),
correlated variability of the fluxes at the low- and high-energy peaks
with no lags is expected (see \cite{vulcano} and references
therein). In blue blazars, the TeV emission is largely due to
lower-energy electrons scattering low-energy (IR) photons, with the
higher-energy electrons ($\gamma > \gamma_{peak}$) producing the
harder TeV photons \cite{tavecchio98}. The X-ray light curve should
track the TeV one, with relative amplitude for the flares at the two
energies in a linear relationship \cite{coppi99}. Thus, {\it TeV
blazars probe the spectrum of the emitting particles}. This is amply
demonstrated by the case of Mrk 421 and Mrk 501, the two brightest and
best-studied TeV blazars. 

\subsection*{Mrk 421} 

\begin{figure}
\noindent{\psfig{file=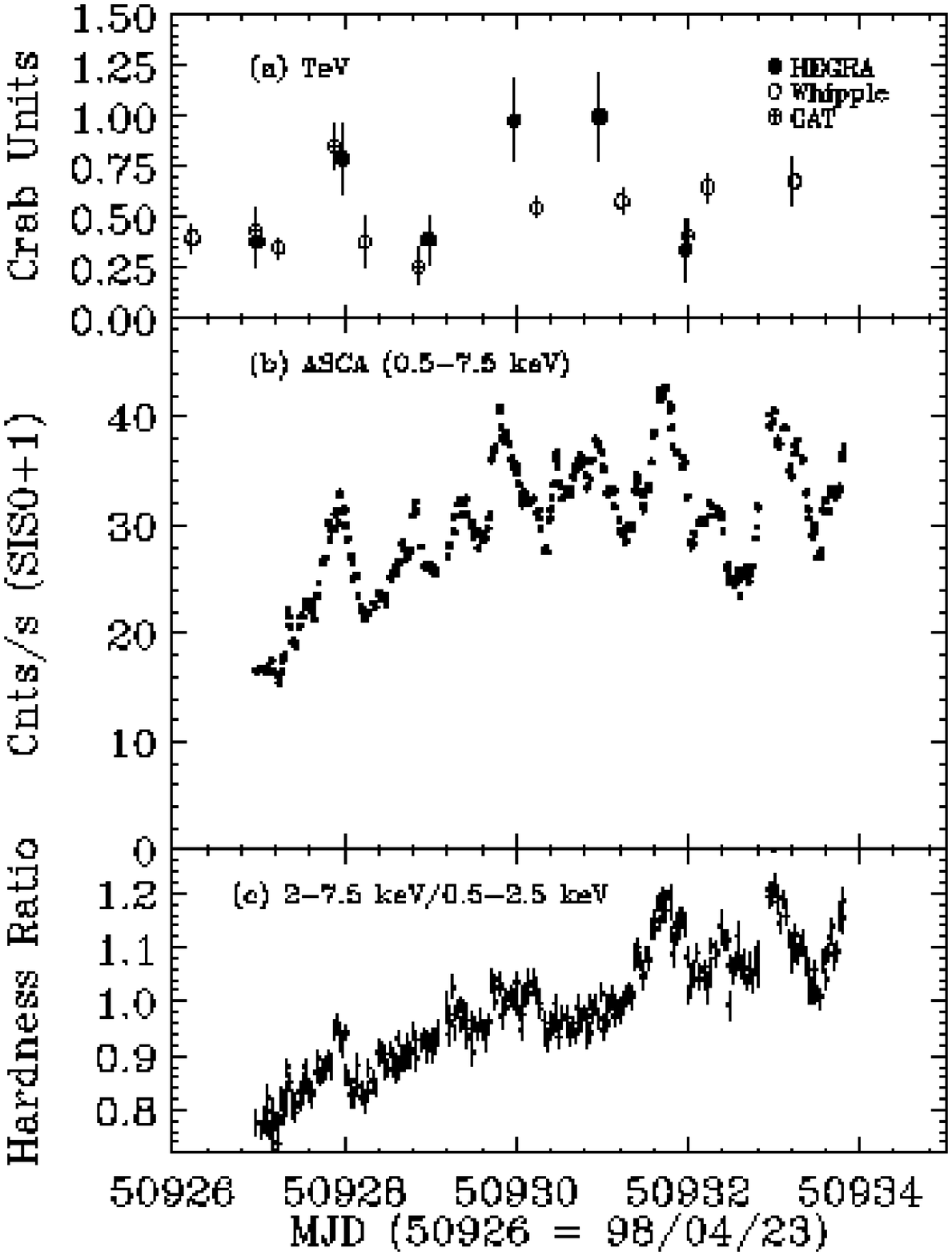,height=3.2in}}{\psfig{file=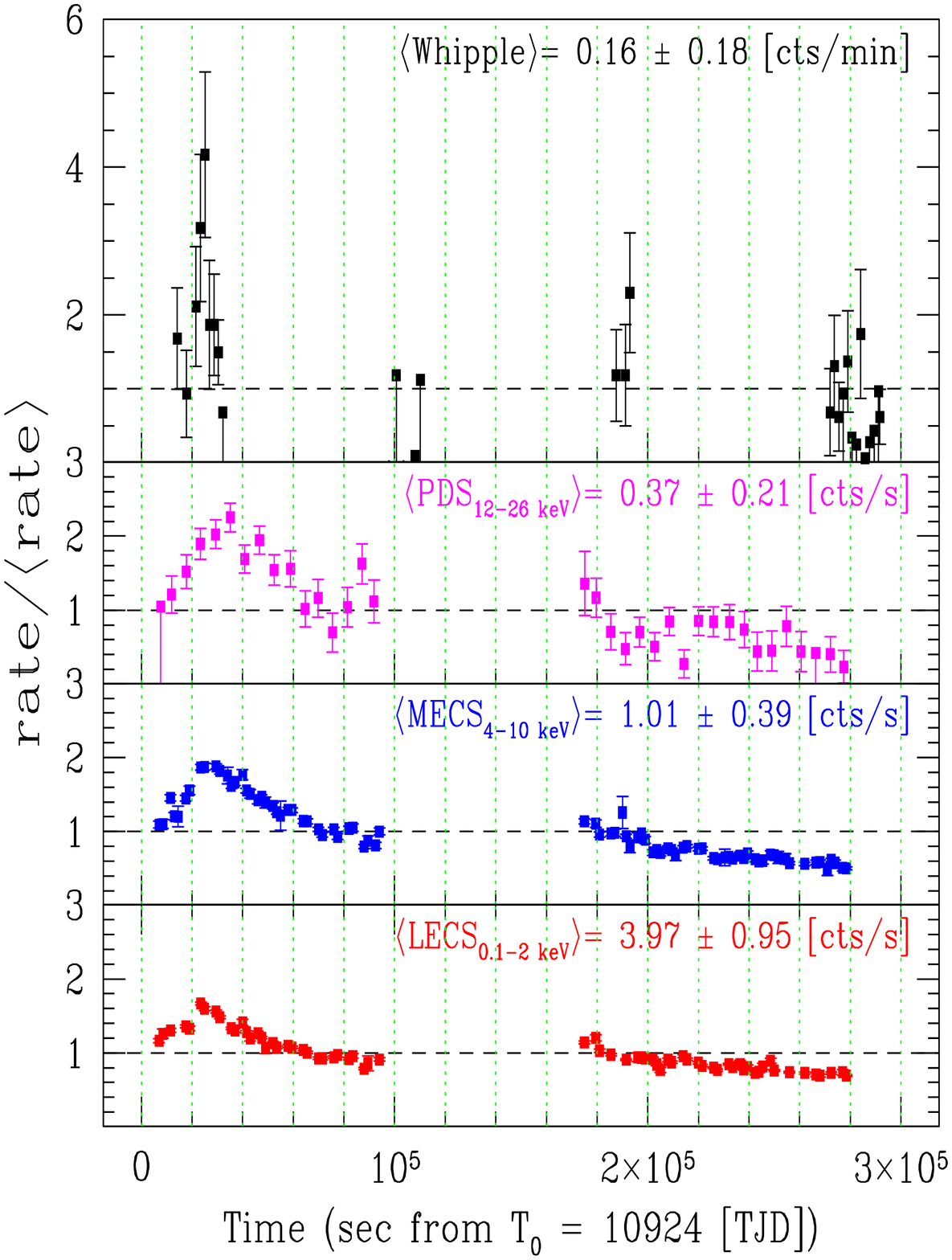,height=3.2in}}
\vspace{10pt}
\caption{Multiwavelength observations of Mrk 421 in 1998 April.
[Left, (a):] Simultaneous TeV and X-ray continuous monitoring over a
period of seven days [20]. Despite the gaps in the sampling, the TeV
light curve tracks the X-ray variations over longer ($\sim$ days)
timescales. Complex flux and spectral variations are observed at
X-rays, with several short flares of $\sim$ one day superposed over a
longer trend and intra-day variations. [Right, (b):] Whipple and SAX
observations of a TeV/X-ray flare at the beginning of the 1998 April
monitoring [21]. The curves are binned at 28 minutes. Correlated X-ray
and TeV flux changes on timescales of hours are detected,
strengthening the idea that the same electron population is
responsible for emitting the X-rays via synchrotron and the TeV
photons via inverse Compton.}
\label{m421_lc}
\end{figure}

Mrk 421 was extensively studied during multifrequency campaigns
conducted in 1994, 1995, and 1998. The early monitorings with ASCA,
Whipple, EUVE, and ground-based telescopes established that the X-ray
and TeV emission is well correlated on longer ($\sim$ days)
timescales, with amplitude generally decreasing with increasing
wavelength \cite{tak96}, although within a rather sparse sampling. An
intensive campaign, continuous over a period of seven days, was
performed in 1998 April, involving ASCA, SAX, EUVE, and various TeV
telescopes \cite{tak99}.  The ASCA and TeV light curves are shown in
Fig. \ref{m421_lc}a.  Complex X-ray flux and spectral variability is
detected by ASCA, with short ($\sim$ 0.5 day) flares superposed on a
longer trend and intra-day variations. The TeV light curve, disrupted
by unfortunate episodes of bad weather, tracks the general trend
observed at X-rays. 

A few days before the start of the ASCA observations and partly
overlapping with it, Mrk 421 was observed with Whipple and SAX, with a
strong flare observed at both wavelengths (Fig. \ref{m421_lc}b). The
new and exciting result is the first detection of X-ray/TeV correlated
variability on timescales of {\it hours} \cite{laura99}, strongly
supporting the idea that the same electron population is responsible
for emitting the X-rays via synchrotron and the TeV photons via
IC. Note the different decay times of the flare, much faster at TeV
than at X-rays (Fig. \ref{m421_lc}b), difficult to explain in the
context of a simple homogeneous model \cite{laura99}. 

\subsection*{Mrk 501}

Mrk 501 attracted much attention in 1997 April when it underwent a
spectacular flare at TeV energies
\cite{catanese97,aharonian99,djannati99}, well correlated with a
similarly-structured X-ray flare detected by RXTE. No delays longer
than one day were detected between the X-ray and TeV emission
\cite{henric99}. A remarkable spectral behavior was observed in the
X-rays (Fig. \ref{seds}b), where an unusually flat (photon index
$\Gamma_X \sim 1.8$) X-ray continuum was measured by SAX and RXTE
during the TeV flare \cite{pian98,henric99}.  This implies a shift of
the synchrotron peak toward higher energies by more than two orders of
magnitude, almost certainly reflecting a large increase of the
electron energy \cite{pian98}, or the injection of a new electron
population on top a quiescent one \cite{kataoka99a}. Later RXTE
observations in 1997 July found the source still in a high and hard
X-ray state \cite{lamer98}, indicating a persistent energizing
mechanism. A similarly flat X-ray continuum was observed in another
weaker TeV blazar, 1ES2344+514 \cite{giommi99}.

\begin{figure}
\noindent{\psfig{file=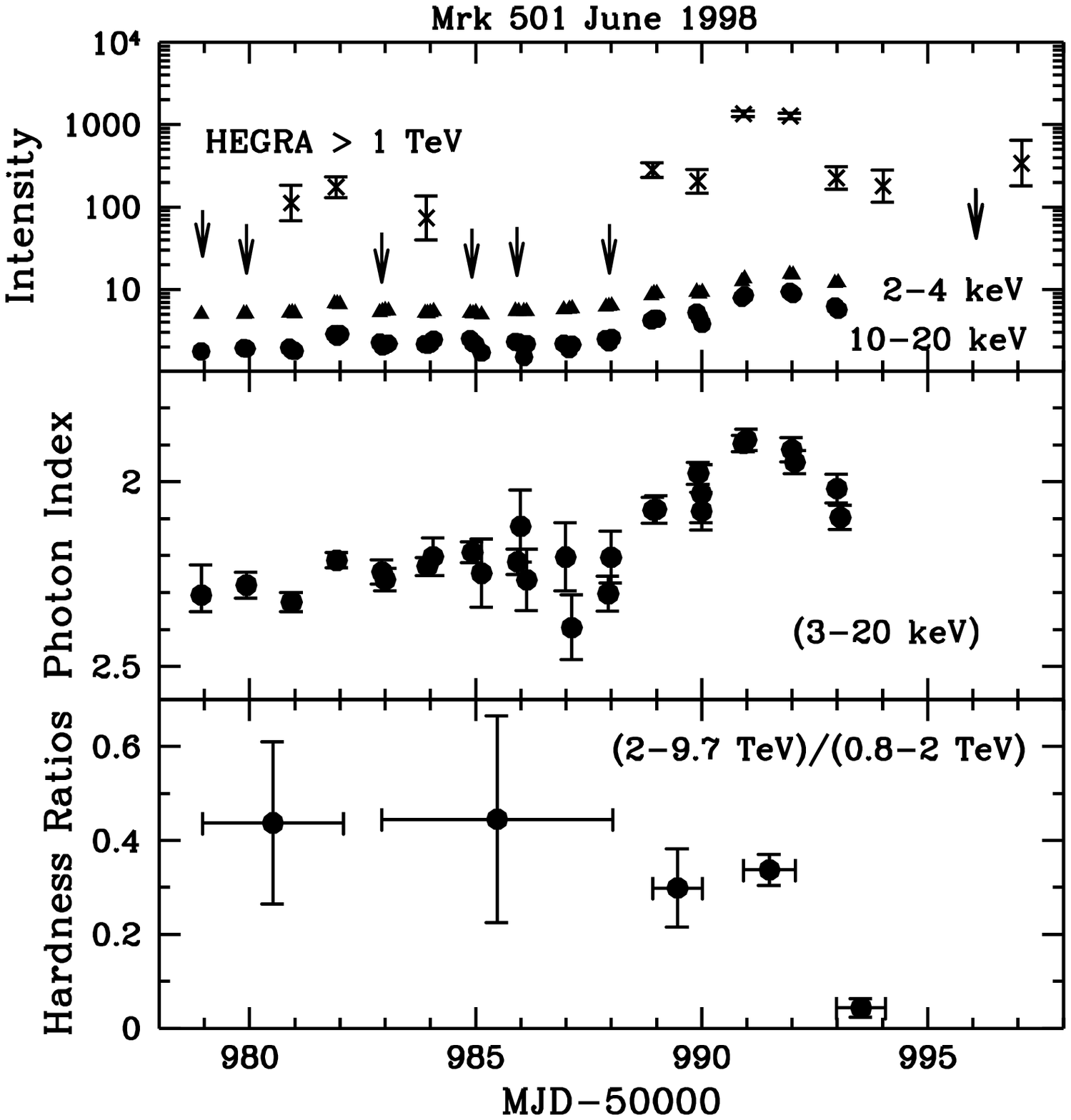,height=3.0in}}{\psfig{file=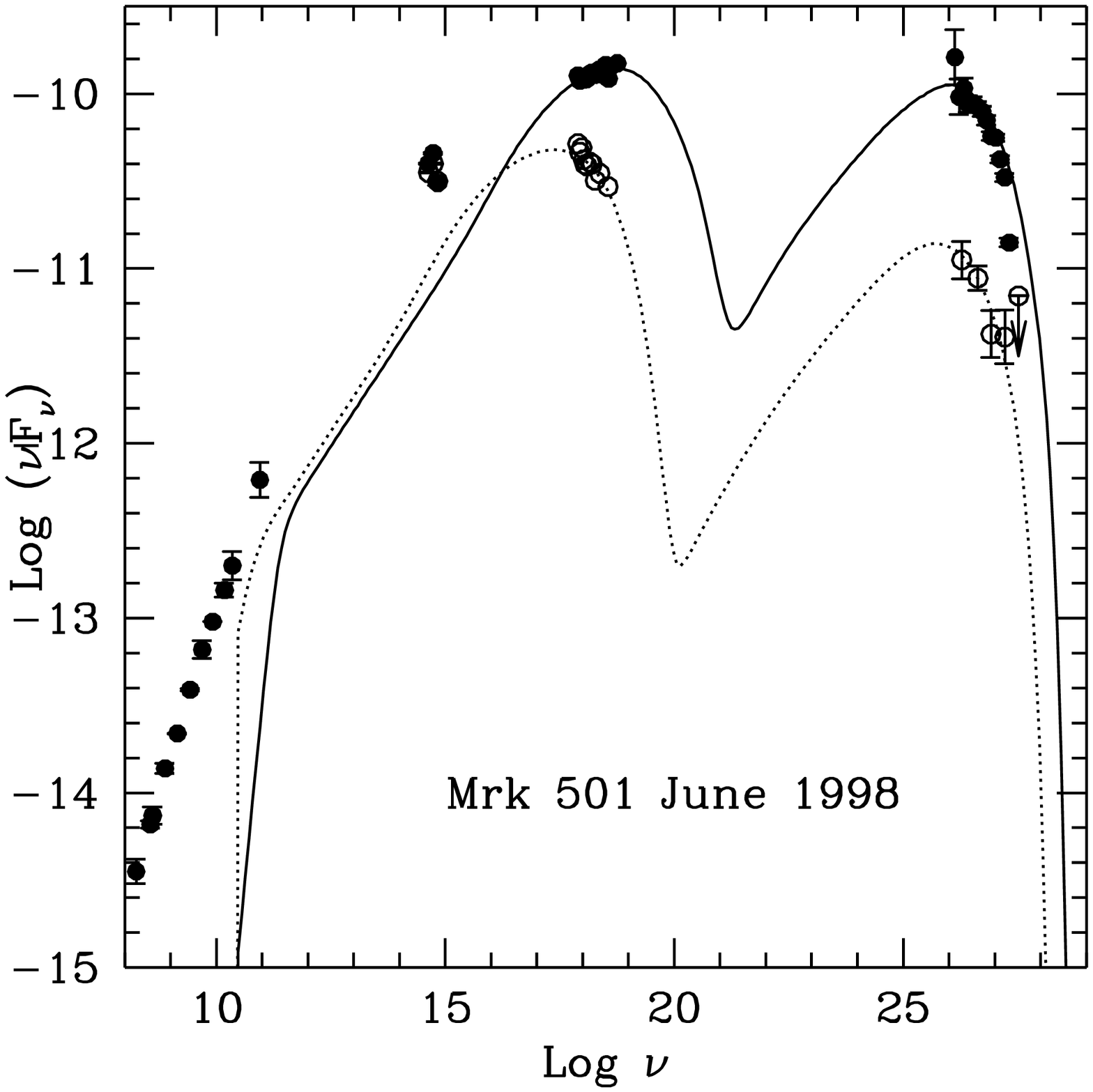,height=2.9in}}
\vspace{10pt}
\caption{Multiwavelength campaign of Mrk 501 in 1998 June with RXTE
and HEGRA [29]. [Left, (a):] TeV and X-ray light curves in different
energy ranges (top panel). A large-amplitude flare is detected at TeV
energies, well-correlated to a shorter-amplitude flare in the
X-rays. Large X-ray spectral variations are observed (middle panel),
while the TeV hardness ratios stay constant during the flare and
soften one day after (bottom panel). [Right, (b):] Spectral energy
distribution during the flare (filled circles) and during quiescence
(open circles), fitted with the SSC model (solid and dotted lines; see
[29] for details).  The large spectral variability at X-rays implies a
shift of the synchrotron peak at $\ge 50$ keV in $\sim$ 2 days, faster
than in 1997. This indicates a similar energizing mechanism at both
epochs operating on different timescales.}
\label{m501-1998}
\end{figure} 

An interesting new behavior was observed during our latest 2-week RXTE
and HEGRA monitoring of Mrk 501 in 1998 June \cite{rita99a}, when
100\% overlap between the X-rays and TeV light curves was achieved
(Fig. \ref{m501-1998}a).  A strong (factor 20 or more), short-lived
($\sim$ two days) TeV flare is detected, well correlated to a
lower-amplitude, broader flare in the X-rays. As in 1997 April, large
X-ray spectral variations are observed in 1998 June, with the X-ray
continuum flattening to $\Gamma_X=1.9$ at the peak of the TeV flare.
This implies a similar shift to $\ge$ 50 keV of the synchrotron peak,
but on much faster timescales (Fig. \ref{m501-1998}b).  However, while
in 1997 April the TeV spectrum hardened during the flare
\cite{djannati99}, as it did in the X-rays, in 1998 June the TeV
hardness ratios stayed relatively constant during the flare and
softened one day later (Fig. \ref{m501-1998}a). The correspondence
between the X-ray and TeV spectra is no longer present during the 1998
June flare.

Intra-hour TeV variability is also detected, with a doubling timescale
for the TeV flux of $\sim$ 30 min \cite{rita99a}. No correlation
with the X-ray light curve on such short timescales was possible, due
to unfortunate gaps in the RXTE sampling. The short TeV variability
timescale implies a size of the emitting region $R \le 5 \times
10^{14}$ cm and a Doppler factor of the emitting plasma $\delta \ge
10$, similar to Mrk 421 \cite{gaidos96}.

\section*{Acceleration and cooling in blazars jets}

Multiwavelength correlated variability of the synchrotron emission
provides tight clues to the structure of the jet through the study of
the energy-dependence of the flare and the accompanying spectral
variations. The rise and decay times of the synchrotron flux depend on
a few source typical timescales \cite{chiaberge99}, while the
accompany spectral variability is a strong diagnostic of the electron
acceleration versus cooling processes \cite{kirk98}, with
characteristic patterns predicted for the hardness ratios as a
function of total intensity depending on the timescales of the two
processes. When cooling dominates, the radiative time can be
approximated by the lag between the shorter and longer synchrotron
wavelengths, providing an estimate of the magnetic field $B$ of the
source via $t_{lag} \sim t_{cool} \propto E^{-0.5} \delta^{-0.5}
B^{-1.5}$, where $E$ is in keV, $B$ in Gauss, and $\delta$ is the
Doppler factor of the emitting plasma \cite{tak96,meg97}.

Two excellent laboratories to study the energy propagation of the
synchrotron flare are PKS 2155--304 and PKS 2005--489, since they are
bright and rapidly variable at all observed wavelengths from optical
to X-rays. They were the targets of recent monitorings, as described
below. 

\subsection*{Multiwavelength observations of PKS 2155--304} 

PKS 2155--304 was detected at TeV energies by the Durham group
\cite{chadwick99} in 1997 November, during a period of intense X-ray
activity \cite{chiappetti99,chadwick99b}, although the source was not
bright enough to allow a detailed TeV light curve. The synchrotron
peak in the SED, usually in the EUV/soft X-ray energy range, shifted
forward one order of magnitude \cite{bertone99}, indicating a more
modest acceleration event than in Mrk 501. Because of its intermediate
SED, the correlated TeV and X-ray variability properties of PKS
2155--304 could be different than the Markarian objects, and this
source qualifies as a high-priority candidate for future X-ray/TeV
monitoring campaigns.

In 1991 November, PKS 2155--304 was observed with 4.5-day continuous
monitoring from optical to UV and X-rays. It exhibited small-amplitude
(10\%), energy-{\it independent} variability, with well-correlated
flares from optical to X-rays on timescales of a few hours and the
shorter wavelengths leading the longer ones \cite{edelson95}. However,
in a subsequent campaign in 1994 May, PKS 2155--304 showed a
substantially different behavior. The source was observed continuously
for $\sim$ 10 days in UV and EUV, and for 2 days in the X-rays and
optical {\cite{meg97}, exhibiting energy-{\it dependent} variations.
A well-defined X-ray flare was observed, followed by broader,
lower-amplitude flares at EUV and UV by $\sim$ 1 and 2 days,
respectively. In X-rays, the harder energies lead the softer ones by
one hour \cite{kataoka99b,zhang99}, implying a magnetic field of $B
\sim 0.1$ Gauss for $\delta=10$, similar to Mrk 421
\cite{tak96,meg97}. This apparent progression of the X-ray flare to
longer wavelengths in 1994 May was explained well by an acceleration
(or equivalent) event in the context of synchrotron radiation, with
the time delay reflecting either synchrotron loss timescales or
physical inhomogeneities of the emission region \cite{meg97}. However,
the quantitatively different behaviors between the campaigns in 1991
and 1994 show that the variability properties of blazars are complex,
involving different modes and likely reflecting an underlying
complexity in the jet structure and/or in the emission mechanisms.

\begin{figure}
\noindent{\psfig{file=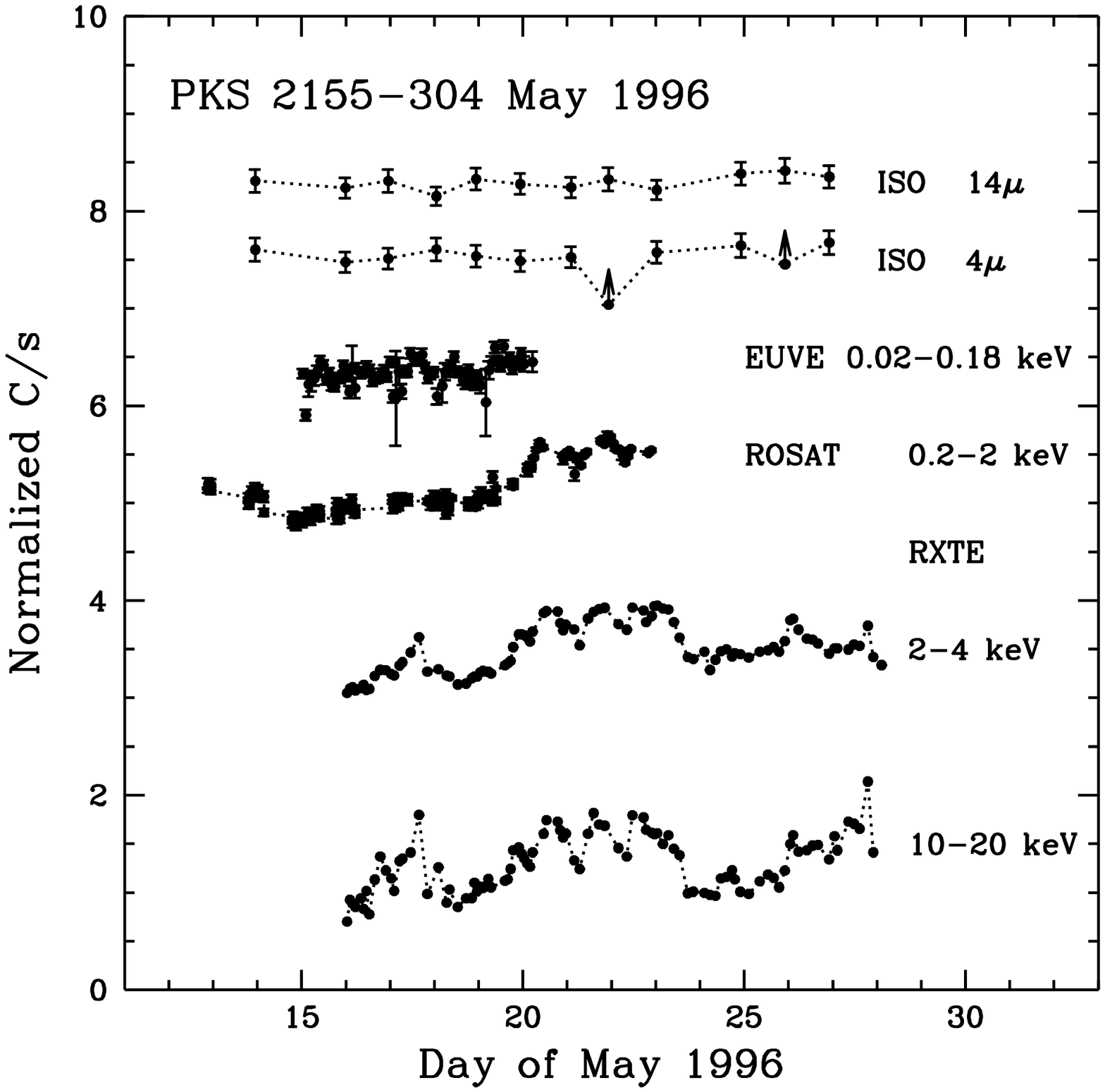,height=2.7in}}{\psfig{file=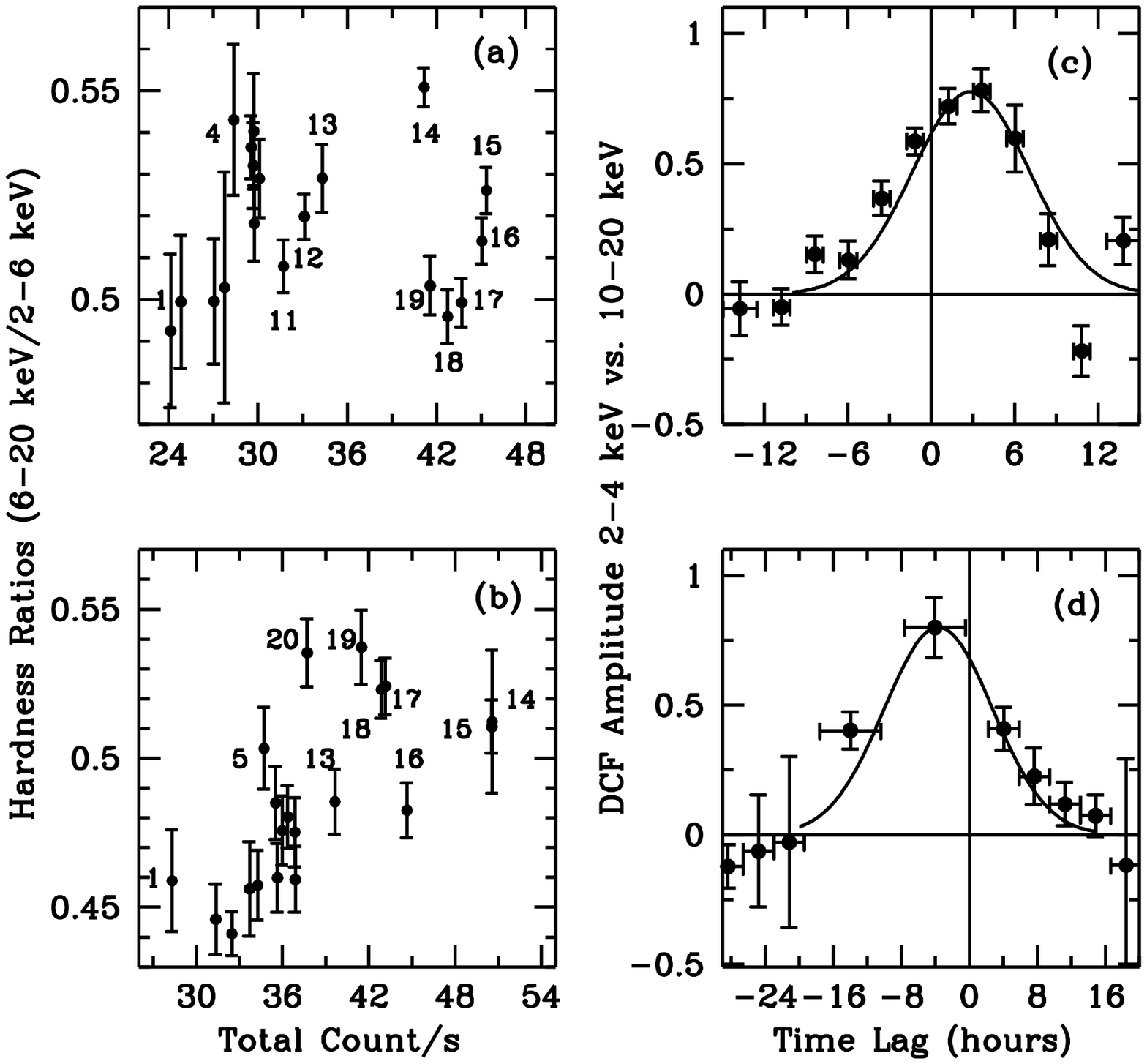,height=2.8in}}
\vspace{10pt}
\caption{Multiwavelength monitoring of PKS 2155--304 in 1996 May
[41]. [Left, (a):] Multiwavelength light curves, showing the excellent
sampling at X-rays and the poorer coverage at the longer wavelengths.
Complex flux and spectral variability was detected in the X-rays, with
energy-dependent amplitude and different flares exhibiting different
spectral behaviors in the hardness ratio versus flux
diagram. Hysteresis loops of both ``clockwise'' and ``anti-clockwise''
signs are observed. [Middle, (b):] Two examples of clockwise (upper
panel) and anti-clockwise (lower panel) loops for the May 18.5--20.2
and 24.2--26.9 flares, respectively.  [Right, (c):] Discrete
Correlation Function applied to the same flares. Soft and hard lags
are detected for the clockwise and anti-clockwise loops, respectively,
of $\sim$ a few hours. This complex spectral behavior is a powerful
diagnostic of the acceleration and cooling processes in the jet.}
\label{2155-lc} 
\end{figure}

PKS 2155--304 was monitored again in 1996 May from IR to X-rays
(Fig. \ref{2155-lc}a; from \cite{rita99b}). In the X-rays excellent
sampling was achieved with RXTE, and a complex flux and spectral
behavior was observed, with short ($\sim$ 1-2 days), energy-dependent
flares superposed to a longer-timescale trend.  Inspection of the
X-ray hardness ratios versus flux shows that the individual flares
exhibit different spectral variability patterns.  Hysteresis loops of
opposite signs, both in a ``clockwise'' (C) and ``anti-clockwise'' (A)
sense, are observed (Fig. \ref{2155-lc}b,c). Analysis of individual
flares with various correlation methods shows that C loops correspond
to soft lags (hard energies varying first) while A loops correspond to
hard lags (soft energies varying first).

This behavior is consistent with a model where energetic electrons are
injected in the source via a shock and escape into the emission
region, where they cool \cite{kirk98}. When acceleration is faster
than cooling, the latter dominates variability and, because of its
energy dependence, the harder energies are emitted first, with C
loops/soft lags observed.  If instead the acceleration is slower
($t_{acc} \sim t_{cool}$), the electrons need to work their way up in
energy and the softer energies are emitted first, yielding A loops and
hard lags. For the first time, we are observing electron acceleration,
which, together with cooling, is responsible for the observed X-ray
variability properties of PKS 2155--304 in 1996 May.

\begin{figure}
\noindent{\psfig{file=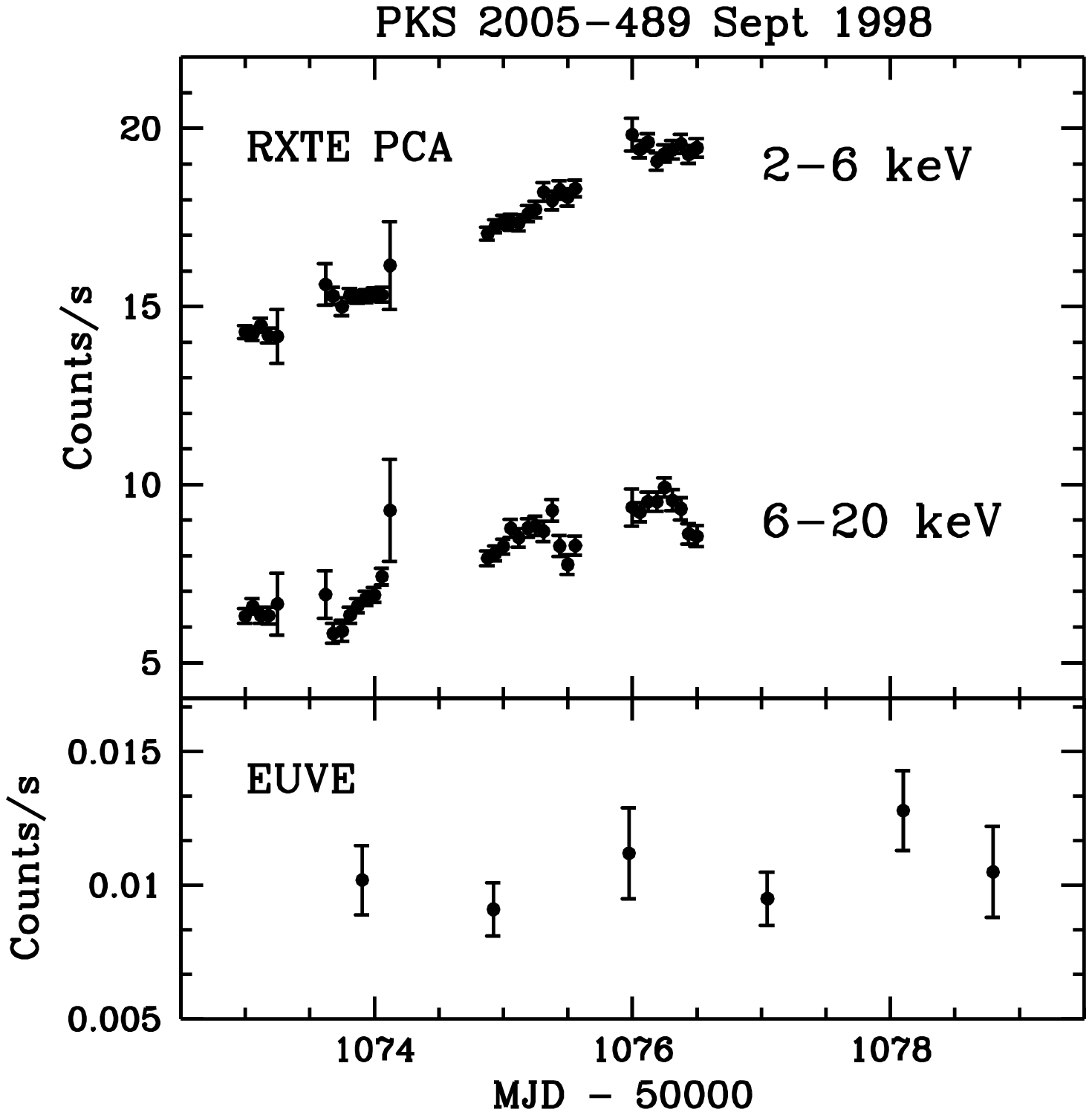,height=3.0in}}{\psfig{file=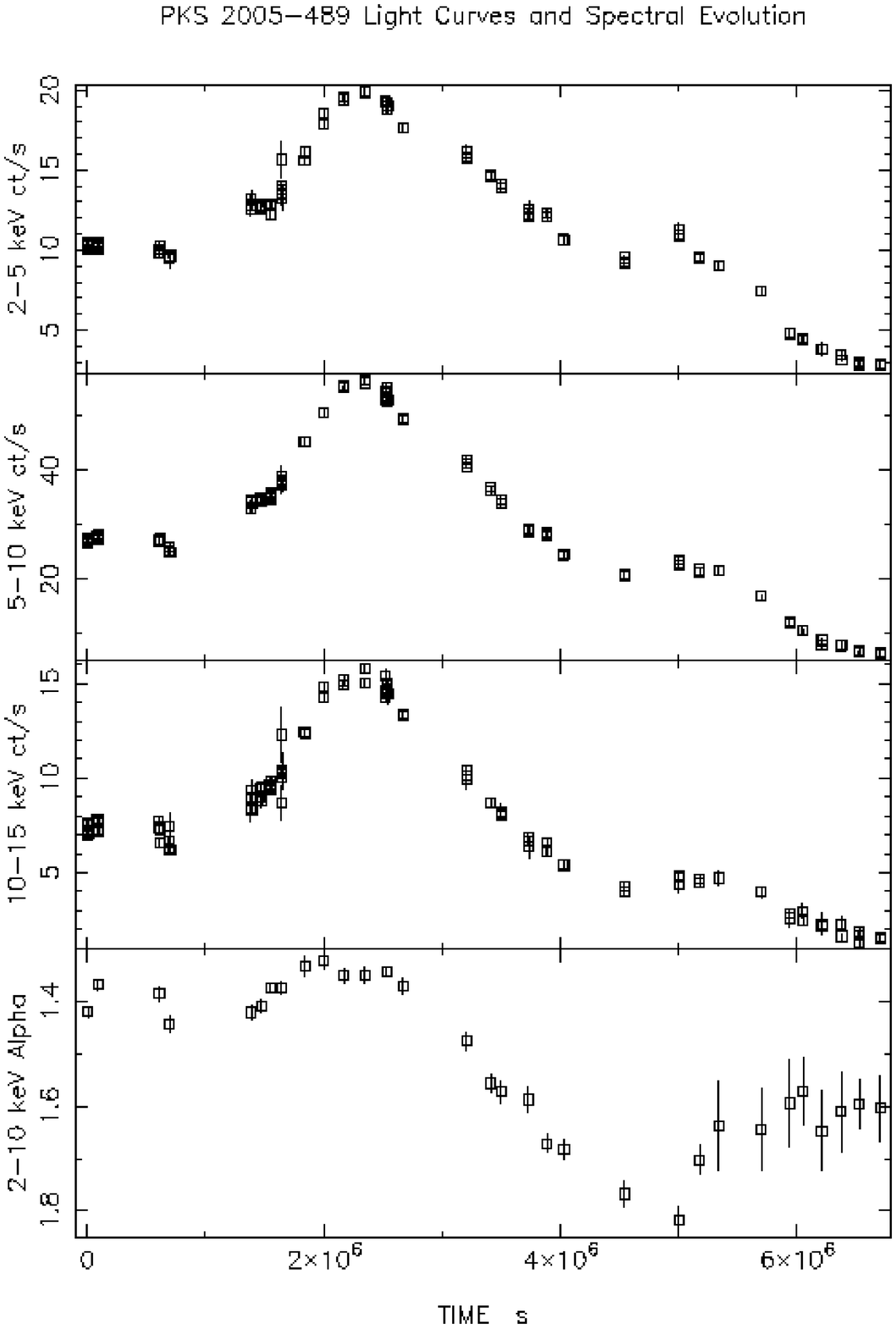,height=3.0in}}
\vspace{10pt}
\caption{ Multiwavelength observations of the TeV candidate PKS
2005--489. [Left, (a):] Our RXTE and EUVE monitoring in 1998
September.  At EUV energies, the source was too faint and no
variability is observed in the daily-binned light curve.  Despite the
gaps in the RXTE sampling, energy-dependent variability is apparent at
X-rays, with a general flux increase of 30\% or more in amplitude.
The harder energies vary before the softer ones, consistent with
cooling dominating the flux variability. [Right, (b):] One month
later, PKS 2005--489 underwent a strong, long-lasting X-ray flare
which was well sampled by RXTE, exhibiting spectral variations on
timescales of hours [45], with similar variability patterns as in 1998
September.}
\label{2005} 
\end{figure}

\subsection*{The TeV candidate PKS 2005--489} 

PKS 2005--489 qualifies as a TeV candidate because of its SED, similar
to Mrk 421, and its proximity ($z$=0.071) \cite{rita95,stecker}.
Observations by the CANGAROO group yielded only an upper limit to the
TeV flux \cite{roberts}.  We monitored this source in 1998 September
with RXTE and EUVE, to study the energy-dependence of the synchrotron
flares. Unfortunately, the source was faint at EUV wavelengths, with a
count rate of $\sim$ 0.01 c/s; no variability is observed in the
daily-binned light curve (Fig. \ref{2005}a).  The X-ray light curves
in a soft (2--6 keV) and hard (6--20 keV) energy band are shown in
Fig. \ref{2005}a. Despite the gaps, due to the RXTE observational
constraints, it is apparent that the variability at hard X-rays is
faster than at soft X-rays, consistent with cooling dominating the
synchrotron flux changes. This is confirmed by the analysis of the
hardness ratios versus flux, where only clockwise loops are
observed. A similar behavior was also observed one month later during
a larger-amplitude, longer-lasting flare, when spectral variations
occurred on timescales of a few hours \cite{perlman99lett}
(Fig. \ref{2005}b).

\begin{figure}
\centerline{\psfig{file=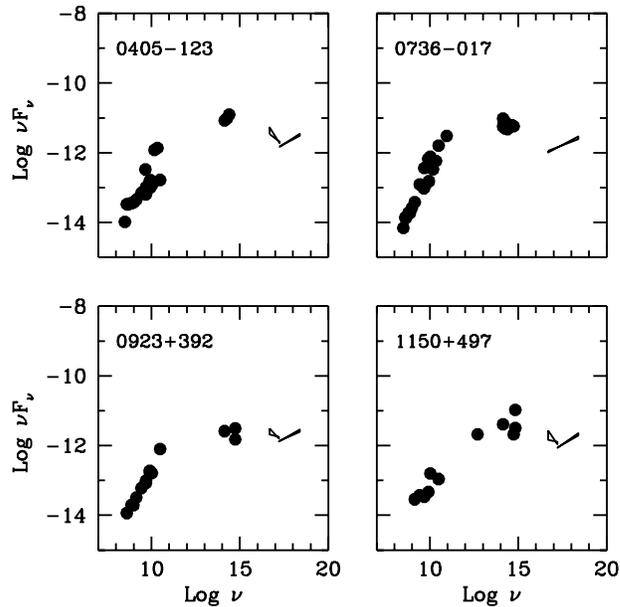,height=3.3in}}
\vspace{10pt}
\caption{Spectral energy distributions of four blue quasars, from our
recent ASCA observations and archival ROSAT and longer-wavelength data
[48].  The ASCA continua are flat, implying an upturn at energies $>$
2 keV. The nature of the optical-to-soft X-ray emission is not well
constrained, and both a thermal (from the accretion disk) and
non-thermal (synchrotron emission from the jet) origin is possible. It
is important to discriminate between these two scenarios with future
observations, since a non-thermal origin would represent a challenge
for current blazar unification schemes.}
\label{fsrq} 
\end{figure}

\section*{Blue Quasars: \\
Undermining the blazar paradigm?} 

A simple and yet powerful probe of the blazar paradigm described above
is provided by the X-ray spectra of the various blazar classes. In
blue blazars, where the synchrotron peak falls at high frequencies,
the X-rays are dominated by the high-energy tail of the synchrotron
component and their X-ray continua should thus be steep and convex, as
a result of radiative losses. On the other end of the spectral
sequence, FSRQs are dominated in the X-rays by the emerging Compton
component and their X-ray continua should be flatter. This is
confirmed by observations of large samples of sources with ROSAT and
ASCA \cite{rita97,kubo}, which yield flat (photon index $\Gamma_X \sim
1.5$) X-ray continua for FSRQs, steep ($\Gamma_X \sim 2.5$) and
downward-curved continua for HBLs, and intermediate slopes for LBLs
($\Gamma_X \sim 2.0$).

It was thus surprising when a sub-group of FSRQs, with otherwise
``canonical'' properties, was observed to have unusually steep ROSAT
spectra, $\Gamma_{0.1-2.4~keV} >$ 2, similar to HBLs
\cite{rita97,padovani96}. These objects were dubbed ``blue quasars''
to indicate that they could be the quasar counterparts of HBLs,
contrary to the predictions of the blazar paradigm which purports that
blue blazars are essentially line-less. Recent deep multicolor
surveys, including our own \cite{perlman99}, are finding an increasing
number of these sources, which now amount to a non-negligible fraction
of the total blazar population. 

What is the true nature of blue quasars? A first simple test is to
measure their hard X-ray continua.  If synchrotron dominates the
optical through X-ray emission, as in HBLs, the X-ray emission above 2
keV should be as steep or even steeper than at softer energies. We
performed exploratory ASCA observations of four blue quasars from our
ROSAT sample \cite{rita97}, selected because relatively nearby ($z <
1$) and {\it not} yet detected at gamma-rays, thus avoiding an {\it a
priori} bias toward flat X-ray slopes. We find that their ASCA spectra
are consistent with flat X-ray continua above 2 keV,
$\Gamma_{2-10~keV} \sim 1.5$, similar to the ``canonical'' FSRQs of
the red type and implying an upturn in the SEDs (Fig. \ref{fsrq}),
most likely the onset of the Compton tail \cite{fsrq99}.

In Fig. \ref{fsrq}, the nature of the emission below 1 keV is not well
constrained, and both a thermal and non-thermal origin is possible. To
this regard, blue quasars could be similar to 3C 273 and 3C 345, two
quasar-like blazars where a thermal ``blue bump'' from the accretion
disk is notoriously present, intruding into the soft X-rays (see
\cite{rita97} and references therein). It is thus entirely plausible
that the large $\alpha_{rx}$ observed in blue blazars (or at least
some of them) is due to a strong thermal contribution from the
accretion disk, as in 3C 273 and 3C 345.  Future simultaneous
optical-UV-X-ray spectra will have the potential to better constrain
the nature of blue quasars and their role in the blazar family.

\section*{Conclusions and Future prospects}

Recent multiwavelength campaigns of blazars expanded the currently
available database, from which we are learning important new lessons.
Detailed modeling of the SEDs of bright gamma-ray blazars of the red
and blue types tend to support the current cooling paradigm, where the
different blazars flavors are related to the predominant cooling
mechanisms of the electrons at the higher energies (EC in more
luminous red sources, SSC in lower-luminosity blue ones).  Future
larger statistical samples are needed, to fully address the
observational biases, especially in gamma-rays.  In particular, it
will be important to expand the sample of TeV blazars, which currently
includes only a handful of sources. 

Correlated multiwavelength variability is the key to understanding the
structure of blazars' jets. In TeV blazars, the X-rays are well
correlated to the TeV emission down to timescales of days and hours
(in Mrk 421), supporting a model where the same electrons are
responsible for the emission at both wavelengths.  It will be
important to determine the shortest timescales on which this
correlation holds, to pin down the electron energy distribution and
the location of the emitting region(s) in the jet. This awaits
well-sampled gamma-ray light curves, which will be afforded by the
next higher-sensitivity missions (HESS, VERITAS, MAGIC, CANGAROO II at
TeV and GLAST at GeV). Broader-band, higher quality gamma-ray spectra
will also be available, allowing a more precise location of the
Compton peak.

An outstanding still unanswered question is the jet composition
(electrons/positrons versus protons). Single-epoch SEDs of blazars are
adequately modeled by both the leptonic and hadronic models, with
different tuning of the parameters. However, while the leptonic models
make specific predictions for the correlated variability properties,
more extensive modeling is currently needed in the context of the
proton jet models. A first effort was presented at this meeting
\cite{rachen99}. 

Finally, coordinated X-ray and longer-wavelength observations of the
synchrotron component in blue blazars strongly suggests that
acceleration, cooling, and escape are the dominant mechanisms
responsible for the observed variability properties.  The case of PKS
2155--304 shows that different flaring modes could be present in a
single source, stressing the importance of multi-epoch monitorings to
obtain a complete picture of the physical processes in blazar jets.

\vspace{1cm} 

This work was funded through NASA contract NAS--38252 and NASA grant
NAG5--7276. I thank Felix Aharonian and the HEGRA team for allowing me
to show the 1998 TeV data of Mrk 501, Eric Feigelson for comments, and
Lester Chou for help with the RXTE data reduction.

\end{document}